\documentclass[pre,onecolumn,amsmath,amssymb,a4paper,superscriptaddress]{revtex4}

\usepackage{amsmath,amssymb}
\usepackage{bm}
\usepackage{graphicx}% Include figure files
\usepackage{dcolumn}% Align table columns on decimal point
\usepackage{bm}% bold math%

\begin{document}

\title{ 
Deformation of a self-propelled domain \\
in an excitable reaction-diffusion system\\
}
\author{Takao Ohta}
\author{Takahiro Ohkuma }
\author{Kyohei Shitara}
\affiliation{ Department of Physics, 
%Graduate School of Science, 
Kyoto University, Kyoto 606-8502, Japan}

\begin{abstract}

We formulate the theory for  a self-propelled domain 
in an excitable reaction-diffusion system in two dimensions where the domain
deforms from a circular shape when the propagation velocity is increased. 
In the singular limit where the width of the domain 
boundary is infinitesimally thin, we derive a set of equations of motion for the center of gravity and 
two fundamental deformation modes. 
The deformed shapes of a steadily propagating domain are obtained.  The set of time-evolution equations exhibits a bifurcation from a straight motion to a circular motion by changing the system parameters.

\end{abstract}

%\pacs{05.45.-a, 82.40.Ck}

\maketitle

%\clearpage

\section{Introduction}

　Self-organized dynamics of domains in reaction-diffusion media have attracted much attention recently \cite{Kapralbook, Swinneybook, PismenBook}. Formation and propagation of domains in excitable reaction-diffusion systems have been studied both theoretically and experimentally.
Not only extended domain patterns such as spiral wave but also interacting disconnected domains have been found in experiments of chemical reactions  where fascinating dynamics due to propagation, collision and self-replication of domains have been observed \cite{Swinney1, Swinney2}.
Throughout the present paper, we shall use the word "domain" for a localized isolated object in two dimensions rather than the word "pulse" which is the concept in one dimension.

 Computer simulations of reaction-diffusion equations have revealed various interesting dynamics of domains \cite{Pearson, Reynolds1, Reynolds2, Petrov, Mikhailov}.  Krisher and Mikhailov have investigated numerically collisions of a pair of domains in two dimensions in an excitable reaction-diffusion system with a global coupling  \cite{Mikhailov}. 
They have also shown that an isolated  domain is deformed substantially from a circular shape when the propagating velocity is increased. It has been  shown that Sierpinski gasket patterns appear in space-time plot of the pulse trajectory in one dimension in several reaction diffusion systems due to a delicate interplay between pair-annihilation and self-replication of pulses upon collision \cite{Hayase, Hayase2, Hayase3}.

It has been known in excitable reaction-diffusion equations that a motionless localized domain loses its stability and begin to propagate when a system parameter exceeds the critical threshold.  Reduction theories in the form of ordinary differential equations in terms of a few degrees of freedom have been developed near this bifurcation called a drift bifurcation not only in one dimension  \cite{Ohta97, Ei} but also  in two and three dimensions \cite{Schenk, Or-Guil, Ohta01, Nishiura05, Pismen01}.
For example, the motion of the center of gravity of an isolated spherical domain obeys in the vicinity of a supercritical drift bifurcation
\begin{equation}
\frac {d \bm{v}}{d  t}
=\gamma \bm{v}- |\bm{v}|^2\bm{v},
\label{eq:eqv}
\end{equation}
where $\bm{v}$ is the velocity. The coefficient $\gamma$ is negative below the threshold whereas it is positive above the threshold. The interaction between two domains can be incorporated in Eq. (\ref{eq:eqv}) by assuming that the characteristic length of the domain is much smaller than the distance between the domains. Here it is remarked that Eq.  (\ref{eq:eqv}) for  $\gamma>0$ is a model equation for the active Brownian motion if a noise term is added \cite{Schweitzer, Condat, Ebelingbook}.
 
 As mentioned above, a domain is generally deformed when the propagating velocity is increased. However, to our knowledge, no analytical theory has been available for dynamics of deformable self-propelled domains. Only a stationary deformed shape of a steadily propagating domain at a constant speed  has been studied by means of the interface dynamics in reaction diffusion systems \cite{Pismen01,  Zykov08}. (In Ref. \cite{Pismen01}, the author claims that a stable propagating domain  exists in a set of two-component reaction diffusion equations whereas the author in Ref. \cite{Zykov08} studies a traveling domain in the slightly different reaction diffusion equations introducing a feedback control term to stabilize the traveling domain.) 
 
  In the present paper, we shall derive the time-evolution equations for the motion of a deformed domain starting with the excitable reaction-diffusion equations studied numerically by Krisher and Mikhailov \cite{Mikhailov}.
 In a previous paper \cite{Ohta09}, we introduced phenomenologically a set of evolution equations for $\bm{v}$ and $S$, the latter of which is the so-called nematic order parameter tensor \cite{Prost} to represent an elliptical deformation of domain.  
  A remarkable property of this system is that there is a bifurcation such that a straight motion of a domain becomes unstable for sufficiently small values of the relaxation rate of $S$ and the domain undergoes a circular motion. Therefore a deformable domain exhibits two bifurcations. One is the drift bifurcation and the other is the straight-circular-motion bifurcation. Hereafter, we shall call this bifurcation a rotation bifurcation.
 
 We shall employ the following assumptions in the derivation of the time-evolution equations for a deformable self propelled particle. First of all, we  represent the domain dynamics by means of  the interface equation of motion for the domain boundary assuming that the interface width is much smaller than any other lengths of the problem \cite{PismenBook}. The second assumption is slowness of  the propagating velocity and hence  weakness of deformation provided that the system is near the drift bifurcation point. 
 %The third assumption is that the higher order time-derivatives of the kinetic variables are ignored, which is justified in the vicinity of the supercritical rotation bifurcation where the angular frequency of rotation or the acceleration of domain is small.
 
 Although the present formulation is specific to the excitable reaction-diffusion equations, the set of the time-evolution equations takes a general form in terms of the vector $\bm{v}$, the second rank tensor $S$ and the third rank tensor $U$. The tensor $U$ was not considered in ref. \cite{Ohta09} but is necessary to represent the head-tail asymmetry of a propagating domain. It is emphasized that the new tensor variable $U$ does not enter up to the leading order in the equations for  $\bm{v}$ and $S$ and hence the previous analysis in ref. \cite{Ohta09} is unaltered.

 We summarize some of the related studies.  First of all, it should be remarked here that there are several experiments of spontaneously deformed propagating domains. For example, an oil droplet in surfactant solutions undergoes a self-propelled motion and changes its shape depending on the propagating velocity \cite{Sumino, Nagai}. The present theory would provide some insight into these experiments.
Next, Shimoyama et al \cite{Sano} have introduced a model of interacting elliptical particles, where the direction of the long axis is chosen to be an independent dynamical variable coupled to the velocity of the center of gravity.   The reaction driven propulsion has been studied both experimentally and theoretically  \cite{Sumino, Nagai, Nagayama, Baer, Golestanian, Pismen06, Kapral}. The active Brownian particles with a harmonic interaction between a pair of particles show a transition from a translational motion to a rotational motion by increasing the noise strength  \cite{Mikhailov05}.
The Langevin equations for the center-of-mass position of rods and the rod orientation have been introduced to study the Brownian circular swimmer in a confined geometry \cite{Loewen}. Collective dynamics of self-propelled particles with an interaction between chemotactic materials and an internal degree of freedom has been investigated  \cite{Tanaka}. However, all of these theoretical studies deal with undeformable objects.

 The present paper is organized as follows. In the next section (section 2), we start with the reaction-diffusion equations for an activator and an inhibitor with a global coupling. In section 3, we briefly review the interface dynamics. The representation of a deformed domain is given in section 4. Apart from the center of gravity, we consider two tensor variables $S$ and $U$ corresponding, respectively,  to the $n=\pm2$, and $\pm3$  modes in the Fourier series expansion of the deformation $\delta R(\phi, t)=\sum_nc_n(t)e^{i\phi n}$ with $\phi$ the angle to specify the location on the interface.
In section 5, the time-evolution equation for the center of gravity is derived whereas, in section 6, the equations for  $n=\pm2$, and $\pm3$ modes are obtained to complete the set of the equations for a deformable self-propelled domain. 
%Some of the properties of the equations obtained 
The approximations used in the derivation of the equations and the stationary shape of a propagating domain are described in section 7.  Summary and discussion are given in section 8.
%In this paper, our formulation is restricted to two dimensions. However, an extension to three dimensions is straightforward and some expressions in three dimensions are given in Appendix. 

\section{ Excitable reaction diffusion system}

We start with a coupled set of reaction-diffusion equations for an activator $u$
and an inhibitor $v$.
\begin{equation}
\tau\epsilon\frac {\partial u}{\partial t}
=\epsilon^2\nabla^2 u +f\{u, v \}-v
 \;,
\label{eq:pequ}
\end{equation}
\begin{equation}
\frac {\partial v}{\partial t}
=D\nabla^2 v+u -\gamma v
 \;,
\label{eq:peqv}
\end{equation}
where
\begin{equation}
f\{u,v \}=-u+\theta (u-p'\{u,v \})
 \;,
\label{eq:peqf}
\end{equation}
with $\theta(x)=1$ for $x>0$ and
$\theta(x)=0$ for $x<0$. The functional $p'\{u,v \}$ contains a global coupling 
as
\begin{equation}
p'=p+\sigma[\int (u+v) d\bm{r}-W]
 \;,
\label{eq:global}
\end{equation}
where $\sigma$ and $W$ are positive constants, 
$0<p<1/2$ and the integral runs over the whole space. 
The constants 
$\tau$ and $\gamma$ are positive and chosen such that the system is
excitable and that a localized stable 
pulse (domain) solution exists. 
Inside the domain, the variable $u$ is positive surrounded by
the rest state where $u$ and $v$ vanishes asymptotically away from the domain.
The parameter $\epsilon$ is a measure of the width of the domain boundary. Hereafter
we consider the limit $\epsilon \rightarrow 0$.

The set of Eqs. (\ref{eq:pequ}) and (\ref{eq:peqv}) with (\ref{eq:peqf}) and
(\ref{eq:global}) was introduced by Krisher 
and Mikhailov \cite{Mikhailov}. They investigated the collision of domains in 
two dimensions by computer simulations and found a reflection of a pair of colliding domains
when the propagating velocity is sufficiently small. 
%The global coupling condition (\ref{eq:global}) can be replaced by a differential equation allowing the time dependence of $a'$. The resulting three-variable model was also simulated \cite{Kawaguchi}, \cite{Schenk} to obtain the results similar to Ref. \cite{Mikhailov}. Or-Guil et al. have derived the equation of motion for an isolated single domain \cite{Or-Guil}
The reason why the global coupling is necessary in the system 
(\ref{eq:pequ}) and (\ref{eq:peqv}) is as follows.
It has been known that a motionless domain is stable when $\tau$ is
sufficiently large. 
There is a bifurcation such that below a threshold value  a
motionless domain becomes unstable and, when the global coupling is absent, a breathing motion occurs such that  the radius of  domain undergoes a periodic oscillation \cite{Koga, Ohta89, Kerner}. 
On the other hand, it is also known that the drift bifurcation exists for smaller values of $\tau$. Therefore, when the value of $\tau$ is decreased,
one generally encounters the situation that 
the breathing bifurcation occurs before the drift bifurcation.
In order to study the intrinsic property of a propagating domain, one must exclude  the breathing instability. 
This is achieved if one chooses a sufficiently large value of $\sigma$ in 
the global coupling (\ref{eq:global})
so that it becomes $p'=p$ with 
\begin{equation}
\int d\bm{r}(u+v)=W 
 \;,
\label{eq:constraint}
\end{equation}
and $f\{u,v \}$ is no more a functional but is given by
\begin{equation}
f(u)=-u+\theta(u-p)
 \;.
\label{eq:fu}
\end{equation}
In the limit $\epsilon \rightarrow 0$ and $\sigma \rightarrow \infty$ in (\ref{eq:global}), Eq. (\ref{eq:pequ}) becomes
\begin{equation}
-u+\theta(u-p)-v=0
 \;.
\label{eq:pequ0}
\end{equation}
 The location of the domain boundary  (interface) is defined such that $u=p$ and $u > (<) p$ inside (outside) the domain. It is shown from Eq. (\ref{eq:pequ0}) that $u+v=1$ inside the domain whereas $u+v=0$ outside the domain. 
Therefore the constraint (\ref{eq:constraint}) 
means that the volume of an excited domain is independent of
time and hence the breathing motion is prohibited. It is also noted that the bifurcation from a motionless domain to a propagating
domain becomes supercritical in the limit $\sigma \rightarrow \infty$ \cite{Mikhailov}.

Substituting Eq. (\ref{eq:pequ0}) into Eq. (\ref{eq:peqv}) yields
\begin{equation}
 \frac{\partial v}{\partial t} =D\bm{\nabla}^2v+\theta(u-p)-\beta v
  \;,
\label{eq:peqv1}
\end{equation}
where $\beta=1+\gamma$.
The equilibrium solution of a motionless circular domain with radius $R_0$ in two dimensions has been obtained in ref. \cite{Ohta89}. Noting the relation that $\theta(u-p)=\theta(R_0-r)$, the equilibrium solution $v=\bar{v}$ is given from Eq.  (\ref{eq:peqv1}) for $0<r<R_0$ by
\begin{equation}
\bar{v}=\frac{1}{\beta}[1-R_0\kappa K_1(R_0\kappa)I_0(r\kappa)]
 \;,
\label{eq:v021}
\end{equation}
and for $r>R_0$ by
\begin{equation}
\bar{v}=\frac{R_0\kappa}{\beta}I_1(R_0\kappa)K_0(r\kappa)
 \;,
\label{eq:v22}
\end{equation}
where 
\begin{equation}
\kappa=\sqrt{\frac{\beta}{D}}
 \;,
\label{eq:kappa}
\end{equation}
and $I_n$ and $K_n$ are the modified Bessel functions. The equilibrium profile of $\bar{u}$
is given by the relation $\bar{u}=\theta(R_0-r)-\bar{v}$.

From Eqs. (\ref{eq:v021}) and (\ref{eq:v22}), it is clear that the inhibitor $v$ changes gradually in space with the characteristic length $\kappa^{-1}$. On the other hand, the activator $u$ is discontinuous at $r=R_0$ in the limit $\epsilon \to 0$ and hence there is a sharp interface.  In this situation, one can derive the interface equation of motion as shown in the next section.

\section{Interface equation of motion}

%In order to make the present paper self-contained, we describe, in this section, some of the results obtained in Ref. \cite{Ohta89}, which will be used in the subsequent sections.

In order to see the spatial variation of $u$ near $r=R_0$, one must rescale the space coordinate as $r'=r/\epsilon \sim  O(1)$. In this length scale, the spatial variation of $v$ is negligible and the value of $v$ in  
(\ref{eq:pequ}) can be replaced by the value at the interface $v(\bm{r}, t)=w$. 
As mentioned above,  the location of the interface is defined through the condition $u(\bm{r}, t)=p$.
For a given value of $w$,
one can readily obtain, from Eq. (\ref{eq:pequ}), 
the equation of motion for an arbitrary deformed interface \cite{Ohta89}
\begin{equation}
\tau V=\epsilon K +\tau c(w) +L
 \;,
\label{eq:interface}
\end{equation}
where $V$ is the normal component of the velocity directed
from the inside to the outside of a domain and $K$ is the mean curvature defined such that it is
positive when the center of the curvature is outside the excited domain. 
The second term in 
(\ref{eq:interface}) is the velocity for a flat interface and is related with $w$ as
\begin{equation}
\frac{c\tau}{\sqrt{(c\tau)^2+4}}=1-2p-2w
 \;.
\label{eq:velocity}
\end{equation}
The unknown constant $w$ is determined by solving Eq.
(\ref{eq:peqv}) or Eq. (\ref{eq:peqv1}) for a given interface configulation.
The last term $L$ is a Lagrange multiplier for the constraint of the domain area conservation (\ref{eq:constraint}). That is, the constant $L$ is determined by the 
condition
\begin{equation}
\int d\omega V=0
 \; ,
\label{eq:L}
\end{equation}
where $d\omega$ is the infinitesimal length (area in three dimensions) of the interface  and the integral runs over the interface. 

When the motion of the interface is slow compared with the relaxation rate of the inhibitor,
one may deal with the term $\partial v/ \partial t$ in (\ref{eq:peqv1}) as a perturbation so that
the asymptotic solution of (\ref{eq:peqv1}) can be written as 
\begin{equation}
v(\bm{r},t)=G\theta-G^2\frac{\partial \theta}{\partial t}+G^3\frac{\partial^2 \theta}{\partial t^2}
-G^4\frac{\partial^3 \theta}{\partial t^3}+...
 \;,
\label{eq:v01}
\end{equation}
where $G$ is defined through the relation 
\begin{equation}
(-D\nabla^2+\beta)G(\bm{r}-\bm{r}')=\delta(\bm{r}-\bm{r}')
\;,
\label{eq:G}
\end{equation}
and the abbreviation such that
$GA=\int d\bm{r}'G(\bm{r}-\bm{r}')A(\bm{r}')$ has been used.
The inhibitor $v$ given by Eq. (\ref{eq:v01}) at the interface position determines the value of $w$.
Substituting it into Eq. (\ref{eq:interface}) with (\ref{eq:velocity})  completes the closed form of the interface
equation of motion.
As discussed in ref. \cite{Ohta01}, the motion of interface is arbitrarily slow in the vicinity of the supercritical drift bifurcation where the expansion  (\ref{eq:v01}) is justified.

\section{Deformed domain}

We consider a deformed domain with the center of gravity $\bm{\rho}(t)$. Its time-derivative $\dot{\bm{\rho}}$ is given by
\begin{equation}
\dot{\bm{\rho}}\equiv\bm{v}= \frac{1}{\Omega}\int d\omega V(\omega)\bm{R}(\omega)
 \;,
\label{eq:drho}
\end{equation}
where the dot means the time derivative, $\Omega$ is the area (volume in three dimensions) of the domain and
\begin{equation}
\bm{R}(\phi) = R(\phi) \bm{e}_r
 \;,
\end{equation}
with  the radial unit vector  $\bm{e}_r$.  The distance from the center of gravity to the interface is denoted by $R(\phi)$ with the angle $\phi$ with respect to the x axis. We assume that $R(\phi)$ is a single-valued function of $\phi$ for sufficiently weak deformations.
The infinitesimal length $d\omega$ along the interface is related with $d\phi$ as
\begin{equation}
d\omega = \left| \frac{d \bm{R}}{d \phi} \right| d\phi = \sqrt{R^{2} + R^{\prime 2}} d\phi
 \;,
\end{equation}
where the prime indicates the derivative with respect to $\phi$.  
%The configuration of the domain is determined by specifying  the vector
%\begin{equation}
%\bm{\rho} + \bm{R}(\phi)
% \;.
%\end{equation}
The tangential unit vector $\bm{t}$ and the unit normal $\bm{n}$ at a position on the interface are given, respectively, by
\begin{eqnarray}
\bm{t} &=& \frac{1}{\sqrt{R^2 + R^{\prime 2}}} \left( R' \bm{e}_r + R \bm{e}_\phi \right)  \;, \\
\bm{n} &=&  \frac{1}{\sqrt{R^2 + R^{\prime 2}}} \left( R \bm{e}_r - R' \bm{e}_\phi \right)  \;,
\end{eqnarray}
where   $\bm{e}_\phi$ is the azimuthal unit vector.  From these expressions, one obtains
\begin{eqnarray}
K(\phi) &=& -\frac{R^2 + 2 R^{\prime 2} - R'' R}{\sqrt{R^2 + R^{\prime 2}}^{3}}  \;, \\
V(\phi) &=& \bm{v}\cdot \bm{n} + \frac{R dR/dt}{\sqrt{R^2 + R^{\prime 2}}} 
 \;.
\end{eqnarray}

Deformations of a domain around a circular shape with radius $R_0$ are written  as 
\begin{equation}
R(\phi) = R_0 + \delta R(\phi, t) 
 \;,
\label{eq:cn}
\end{equation}
where
\begin{eqnarray}
 \delta R(\phi, t)  = \sum_{n=-\infty}^{\infty} c_n(t) e^{i n \phi}
  \;.
\label{eq:deltaR}
\end{eqnarray}
Note that since the translational motion of the domain has been incorporated in the variable $\bm{\rho}$, the modes $c_{\pm 1}$ should be removed from the expansion (\ref{eq:deltaR}).
The mean curvature and the normal component of the velocity are given up to the first order of the deviations, respectively,  by
\begin{eqnarray}
K(\phi, t) &=& -\frac{1}{R_0} -\frac{1}{R_0^2}\sum_{n= -\infty}^{\infty}(n^2-1) c_n(t) e^{i n \phi}
 \;,
\label{eq:K1} \\
%\end{eqnarray}
%\begin{eqnarray}
V(\phi, t) &=& \bm{v}\cdot \bm{n}  +   \sum_{n= -\infty}^{\infty}\dot{c}_n(t) e^{i n \phi}
\label{eq:V1}
 \;,
\end{eqnarray}

For an isolated domain, the Fourier transform of $\theta(u-p)=\theta(R-|\bm{r}-\bm{\rho}|)$ is given by
\begin{equation}
\theta_q=\int_{r<R} d\bm{r}\exp(-i\bm{q}\cdot\bm{r})
 \;.
\label{eq:thetaq}
\end{equation}
Here the Fourier transformation is defined by
\begin{eqnarray}
A(\bm{r},t)&=&\int_{\bm{q}} A_{\bm{q}}(t) e^{i\bm{q}\cdot{\bm{r}}}\;, 
\label{eq:Fourier1} \\
 A_{\bm{q}}(t) &=&\int d\bm{r} A(\bm{r},t)e^{-i\bm{q}\cdot{\bm{r}}}\;, 
\label{eq:Fourier2} 
\end{eqnarray}
where $\int_{\bm{q}}=\int d\bm{q}/(2\pi)^d$ with $d$ the dimensionality of space. 
Substituting  (\ref{eq:cn}) into  (\ref{eq:thetaq}) yields up to first order of the deviations
\begin{equation}
\theta_q (t) =\theta_q ^{(0)}  + \theta_q ^{(1)} 
 \;,
\end{equation}
where 
\begin{eqnarray}
\theta_q^{(0)}  &=& \frac{2 \pi R_0}{q} J_{1} (q R_0)  \;, \nonumber \\
\theta_q^{(1)} &=& 2\pi R_0 \sum_{n} c_n (t)i^{-n} e^{in \phi}  J_n (q R_0)
 \;,
\label{eq:theta1}
\end{eqnarray}
and 
$J_n (x)$ is the Bessel function of the $n$-th order defined by
\begin{equation}
\int_{0}^{2 \pi} d \theta e^{i(n \theta \pm z \cos\theta)} = 2\pi i^{\pm n} J_{n} (z)
 \;.
\end{equation}
Here we summarize the formulas for the Bessel function
\begin{eqnarray}
J_{-n} (z) &=& (-1)^{n} J_{n} (z)  \;, \\
%\end{equation}
%\begin{equation}
z^{n} J_{n-1} (z)&=&\frac{d}{dz} \left( z^{n} J_{n} (z)  \right)  \;, \\
%\end{equation}
%\begin{eqnarray}
  \frac{2n}{z} J_{n} (z) &=&J_{n-1}(z) + J_{n+1}(z)  \;,  \\
2 J_{n}^{\prime} (z) &=& J_{n-1}(z) - J_{n+1}(z)   \;.
\end{eqnarray}

Since the amplitude $c_n$ is complex, it is convenient to introduce the linear combinations  of its real and imaginary parts.
The modes $c_{\pm2}$ represents an elliptical shape of domain.
We introduce a second rank tensor as follows; 
\begin{eqnarray}
S_{11}&=&-S_{22}=c_2+c_{-2} \;, \nonumber \\
S_{12}&=&S_{21}=i(c_2-c_{-2}) \;.
\label{eq:S}
\end{eqnarray}
For an elliptical domain, $R(\phi)$ is represented as
\begin{eqnarray}
R(\phi)=R_0+\frac{\delta_2}{2}\cos 2(\phi-\phi_2)  \;,
\label{eq:Re2}
\end{eqnarray}
where $\delta_2$ is a positive constant and $\phi_2$ is the angle between the long axis of the elliptical domain and the x-axis. The tensor $S$ can be written in terms of $\phi_2$ as $S_{11}=(\delta_2/2)\cos 2\phi_2$ and $S_{12}=(\delta_2/2)\sin 2\phi_2$. This is represented in terms of  the unit normal $\bm{N}=(\cos \phi_2, \sin \phi_2)$ along the long axis  as
\begin{eqnarray}
S_{\alpha \beta}=\delta_2(N_{\alpha}N_{\beta}-\frac{\delta_{\alpha \beta}}{2}\bm{N}^2)  \;,
\label{eq:tenS}
\end{eqnarray}
which is the same as the nematic order parameter tensor in liquid crystals \cite{Prost}.

The modes $c_{\pm3}$ are necessary to represent the head-tail asymmetry of a propagating domain. If the deformation $R(\phi)$ is written as
\begin{eqnarray}
R(\phi)=R_0+\delta_3\cos 3(\phi-\phi_3)  \;,
\label{eq:Re3}
\end{eqnarray}
with a positive constant $\delta_3$, one obtains 
\begin{eqnarray}
T_1=c_3+c_{-3} =\delta_3 \cos3 \phi_3   \;, \nonumber \\
T_2=i(c_3-c_{-3})=\delta_3 \sin3 \phi_3  \;.
\label{eq:T}
\end{eqnarray}
It is convenient to introduce a third-rank tensor associated with the  $n=\pm 3$ modes as follows;
\begin{eqnarray}
U_{\alpha \beta \gamma}=\frac{4\delta_3}{3}\sum_{m=1,2,3}N_{\alpha}^{(m)}N_{\beta}^{(m)}N_{\gamma}^{(m)}  \;,
\label{eq:tenU}
\end{eqnarray}
where 
\begin{eqnarray}
\bm{N}^{(1)}&=&(\cos \phi_3, \sin \phi_3)  \;, \\
\label{eq:N1}
\bm{N}^{(2)}&=&\big(\cos (\phi_3 +\frac{2\pi}{3}), \sin(\phi_3 +\frac{2\pi}{3}) \big)  \;, \\
\label{eq:N2}
\bm{N}^{(3)}&=&\big(\cos (\phi_3 -\frac{2\pi}{3}), \sin(\phi_3 -\frac{2\pi}{3}) \big)  \;.
\label{eq:N3}
\end{eqnarray}
From these definitions, one obtains
%\begin{eqnarray}
$U_{111}=T_1$ and $
U_{222}=-T_2$.
%\label{eq:U111U222}
%\end{eqnarray}
Furthermore, it is readily shown that there are relations among the components as
\begin{eqnarray}
U_{111}&=&-U_{122}=-U_{212}=-U_{221}  \;,\nonumber \\
U_{222}&=&-U_{112}=-U_{121}=-U_{211} \;.
\label{eq:Ucomp}
\end{eqnarray}
The tensor  (\ref{eq:tenU}) is the same as the order parameter for banana (tetragonal nematic) liquid crystals in two dimensions \cite{Fel, Brand}.
%The representation of a deformed domain in three dimensions is given in Appendix.

In the following sections, we shall derive a coupled set of the time-evolution equations for $\bm{v}$, $c_{\pm2}$ and $c_{\pm3}$. 
However, it is impossible to derive the set of equations exactly in a general condition. As mentioned in Introduction, we employ an assumption, that is, the smallness of the propagating velocity. Since the circular shape of a domain is assumed to be stable when it is motionless, the deformation of the domain is expected to be weak near the drift bifurcation where the propagating velocity is small \cite{Ohta01}.
In this condition, the trancation of the modes up to $n=\pm3$ is justified. %Another assumption is the smallness of the second derivatives $\ddot{\bm{v}}\equiv d\ddot{\bm{\rho}}/dt$, $\ddot{c}_{\pm2}$ and $\ddot{c}_{\pm3}$. In a previous paper \cite{Ohta09}, we have shown that a deformed self-propelled domain exhibits a rotation bifurcation such that a straight motion of a domain becomes unstable and a circular motion appears. In the vicinity of this supercritical bifurcation, the acceleration is expected to be sufficiently small and hence the higher order time derivatives are safely ignored.

%%%%%%%%%%%%%%%%%%%%%%%%%%%%%%%%%%

\section{Equation of motion for $\bm{\rho}(t)$ }

The aim of this section is to derive the equation of motion for the center of
gravity from the interface equation (\ref{eq:interface}). 
When the velocity is small, one may expand (\ref{eq:interface}) with (\ref{eq:velocity}) in powers of $V$.
Up to the third order, one obtains 
%for $\epsilon \to 0$
\begin{equation}
2w+\frac{\tau V-\epsilon K -L}{2} -\frac{(\tau V-\epsilon K-L)^3}{16}=1-2p
 \;.
\label{eq:vvv}
\end{equation}
The value  $w$ of the inhibitor $v$ at the interface is given from Eq.  (\ref{eq:v01}) by
\begin{equation}
w=w_{(0)}+w_{(1)}+w_{(2)}+w_{(3)}
 \;,
\label{eq:omegaw}
\end{equation}
where
\begin{eqnarray}
w_{(0)}&=&\int_{\bm{q}}G_q \theta_q
e^{i\bm{q}\cdot\bm{R}(\omega)}
 \;,
\label{eq:w0q} \\
w_{(1)}&=&w_{(11)}+w_{(12)} \nonumber \\
&=&i\int_{\bm{q}}(\bm{v}\cdot \bm{q})G_q^2 \theta_qe^{i\bm{q}\cdot\bm{R}(\omega)} \nonumber \\
&-&\int_{\bm{q}}G_q^2(\frac{\partial \theta_q}{\partial t})e^{i\bm{q}\cdot\bm{R}(\omega)}
 \;,
\label{eq:w1q} \\
w_{(2)}&=&w_{(21)}+w_{(22)}+w_{(23)}  \nonumber \\
&=&-i\int_{\bm{q}}(\bm{\dot{v}}\cdot \bm{q})G_q^3 \theta_qe^{i\bm{q}\cdot\bm{R}(\omega)} \nonumber \\
&-&\int_{\bm{q}}(\bm{v}\cdot \bm{q})^2G_q^3 \theta_qe^{i\bm{q}\cdot\bm{R}(\omega)} \nonumber \\
&+&\int_{\bm{q}}G_q^3(\frac{\partial^2 \theta_q}{\partial t^2})e^{i\bm{q}\cdot\bm{R}(\omega)} % \nonumber \\
 \;,
\label{eq:w2q} \\
w_{(3)}&=&-i\int_{\bm{q}}(\bm{v}\cdot \bm{q})^3G_q^4 \theta_qe^{i\bm{q}\cdot\bm{R}(\omega)}
 \;,
\label{eq:w3q}
\end{eqnarray}
with
\begin{equation}
G_q=\frac{1}{D(q^2+\kappa^2)}
 \;.
\label{eq:Gq}
\end{equation}
 We have dropped out terms having a factor $\bm{v}\dot\theta$ in (\ref{eq:w2q}) 
and $\bm{v}\bm{v}\dot\theta$ and the second derivatives in (\ref{eq:w3q}).

In order to obtain the equation for $\bm{\rho}$, we operate $(1/\Omega)\int d\omega \bm R(\omega)$ to Eq. (\ref{eq:vvv}). As mentioned at the end of the preceding section, the basic approximation is the expansion in terms of $\bm{v}$ and $c_{n}$ ignoring the higher order time derivatives. The validity of these approximations will be discussed in detail in section 7.

 Since the Lagrange multiplier $L$ is independent of the angle $\phi$, one has  $\int d\omega \bm R(\omega)L=\int d\omega \bm R(\omega)L^3=\int d\omega \bm R(\omega)V^2L=0$ for a circular domain. Furthermore, it will be shown later in section 6 (just after Eq.  (\ref{eq:Gell})) that $L\sim O(\bm{v}^2)$, one may ignore $\int d\omega \bm R(\omega)VL^2$ up to the third order of $\bm{v}$.
It is also noted that $\int d\omega \bm R(\omega)K(\omega)=0$ up to the first order of deformations since the modes $c_{\pm 1}$ are excluded in Eq. (\ref{eq:deltaR}). The term $\epsilon VK$ and the higher order terms of $\epsilon$ are ignored.
As a result, $\epsilon K +L$ in the third term on the left hand side of Eq. (\ref{eq:vvv}) can be dropped out in the derivation of equation for $\bm{\rho}$.

The zero-th order terms in Eq. (\ref{eq:vvv}) which represent a motionless circular domain are given by
\begin{equation}
-\frac{\epsilon}{2R_0}+1-2p-2\int_{\bm{q}}G_q \theta_q^{(0)}e^{i\bm{q}\cdot\bm{R}^{(0)}(\omega)}=0
 \;,
\label{eq:eqR}
\end{equation}
with $\bm{R^{(0)}}=R^{(0)}\bm{e}_r$. The Lagrange multiplier $L$ can be omitted at this order since it is absorbed into the constant $p$.
Equation  (\ref{eq:eqR})  gives us the equilibrium radius $R_0$ of the circular domain as a function, e.g., of 
the parameter $p$ \cite{Ohta89}.
Since the radius has been fixed by the constraint (\ref{eq:constraint}), this implies
that the parameter $p$ is not independent but should be related to $W$ in  (\ref{eq:constraint}). 
%It should also be noted that since we have ignored the curvature term in (\ref{eq:vvv}), the unstable equilibrium solution with a smaller radius is not obtained from (\ref{eq:eqR}) \cite{Ohta89}.

From the higher order terms in Eq. (\ref{eq:vvv}), one obtains in two dimensions
\begin{equation}
2\bm{w}+\frac{\tau}{2}\bm{v}-\frac{3\tau^3}{64}\bm{v}|\bm{v}|^2=0
 \;,
\label{eq:vvv2}
\end{equation}
where
\begin{equation}
\bm{w}=\bm{w}_{(11)}+\bm{w}_{(21)}+\bm{w}_{(3)}
 \;,
\label{eq:bmw}
\end{equation}
with
\begin{eqnarray}
\bm{w}_{(11)}&=&\frac{1}{\Omega}\int d\omega\int_{\bm{q}}(\bm{v}\cdot \bm{q})G_q^2 \theta_q \frac{\partial}{\partial \bm{q}}e^{i\bm{q}\cdot\bm{R}(\omega)}
 \;,  \\
\label{eq:bmw11}
%\end{equation}
%\begin{equation}
\bm{w}_{(21)}&=&-\frac{1}{\Omega}\int d\omega\int_{\bm{q}}(\bm{\dot{v}}\cdot \bm{q})G_q^3 \theta_q \frac{\partial}{\partial \bm{q}}e^{i\bm{q}\cdot\bm{R}(\omega)}
 \;,  \\
\label{eq:bmw21}
%\end{equation}
%\begin{equation}
\bm{w}_{(3)}&=&-\frac{1}{\Omega}\int d\omega\int_{\bm{q}}(\bm{v}\cdot \bm{q})^3G_q^4 \theta_q \frac{\partial}{\partial \bm{q}}e^{i\bm{q}\cdot\bm{R}(\omega)}
 \;.
\label{eq:bmw3}
\end{eqnarray}
These have been evaluated for a circular domain in two dimensions previously \cite{Ohta01}.
%\begin{eqnarray}
%\bm{w}_{(11)}^{(0)}&=&-\frac{1}{2}\bm{v}\int_{\bm{q}}q^2B_qG_q^2\theta_q^{(0)}
 %\;,
%\label{eq:bmw110} \\
%\bm{w}_{(21)}^{(0)}&=&\frac{1}{2}\bm{\dot{v}}\int_{\bm{q}}q^2B_qG_q^3 \theta_q^{(0)}
% \;,
%\label{eq:bmw210} \\
%\bm{w}_{(3)}^{(0)}&=&\frac{3}{8}\bm{v}|\bm{v}|^2\int_{\bm{q}}q^4B_qG_q^4 \theta_q^{(0)}
% \;,
%\label{eq:bmw30}
%\end{eqnarray}
%where the function $B_q$ is defined by
%\begin{equation}
%B_q=-\frac{R_0}{\Omega}\frac{\partial}{q\partial q}\int _0^{2\pi}d\phi e^{i\bm{q}\cdot\bm{R}^{(0)}(\omega)}=\frac{2}{q}J_1(qR_0)
% \;.
%\label{eq:Bq}
%\end{equation}
Substituting those results into Eq. (\ref{eq:vvv2}) gives us
\begin{equation}
m\bm{\dot{v}} +\frac{1}{2}(\tau-\tau_c)\bm{v}  +g\bm{v}|\bm{v}|^2=-2\delta \bm{w}_{(11)}
 \;,
\label{eq:vvv3}
\end{equation}
where 
\begin{eqnarray}
m&=&2R_0\int_{0}^{\infty}dqqG_q^3J_1(qR_0)^2
 \;,
\label{eq:m} \\
\tau_c&=&4R_0\int_{0}^{\infty}dqqG_q^2J_1(qR_0)^2
 \;,
\label{eq:tauc} \\
g&=&\frac{3R_0}{2}\int_{0}^{\infty}dqq^3G_q^4J_1(qR_0)^2-\frac{3\tau^3}{64}
 \;.
\end{eqnarray}
The left hand side of Eq. (\ref{eq:vvv3}) has been obtained in ref. \cite{Ohta01}.
The right hand side is defined as $\delta \bm{w}_{(11)}=\bm{w}_{(11)}-\bm{w}_{(11)}^{(0)}$ and vanishes identically for a circular domain where $\bm{w}_{(11)}^{(0)}$ is given by (\ref{eq:bmw11}) with $\theta_q=\theta_q^{(0)}$ and $\bm{R}(\omega)=\bm{R}^{(0)}(\omega)$. The coefficient $m$ is positive and $g$ is shown to be positive for $\tau \approx \tau_c$. Therefore, Eq. (\ref{eq:vvv3}) indicates a bifurcation (drift bifurcation) such that a motionless circular  domain $\bm{v}=0$ is stable for $\tau >\tau_c$ whereas it looses stability for $\tau <\tau_c$ and undergoes propagation at the velocity $|\bm{v}|^2=(\tau_c-\tau)/(2g)$.

The right hand side $\delta \bm{w}_{(11)}$ of Eq. (\ref{eq:vvv3}) is given up to the order of $O(\bm{v} c_n)$ by
\begin{eqnarray}
\delta \bm{w}_{(11)}&\equiv& \delta \bm{w}_{(11)}^{(1)}+\delta \bm{w}_{(11)}^{(2)} \nonumber \\
&=&\frac{1}{\Omega}\int d\omega \int_{\bm{q}}(\bm{v}\cdot \bm{q})G_q^2 \theta_q^{(1)} \frac{\partial}{\partial \bm{q}}e^{i\bm{q}\cdot\bm{R}^{(0)}(\omega)}  \nonumber \\
&+& \frac{1}{\Omega}\int d\omega \int_{\bm{q}}(\bm{v}\cdot \bm{q})G_q^2 \theta_q^{(0)} \frac{\partial}{\partial \bm{q}}e^{i\bm{q}\cdot\bm{R}^{(0)}(\omega)}
 \;.
\label{eq:dbmw11}
\end{eqnarray}
The following formula is useful to evaluate $\delta \bm{w}_{(11)}$;
\begin{eqnarray}
& &\frac{1}{\Omega}\int d\omega e^{i\bm{q}\cdot\bm{R}(\omega)} =
\frac{2}{R_0}J_0(qR_0) \nonumber \\
&+&\frac{q}{\pi R_0}\sum_{n} c_n (t)e^{in \theta} 2\pi i^n J_{n-1} (q R_0) \nonumber \\
&+&\frac{1}{\pi R_0^2}\sum_{n} c_n (t)e^{in \theta}2\pi i^n (1-n)J_{n} (q R_0)
 \;,
 \label{eq:form}
\end{eqnarray}
where $\theta$ is the angle between $\bm{q}$ and the $x$ axis, i.e., $\bm{q}=q(\cos \theta, \sin \theta)$.
This formula is valid up to the first order of the deformation.

Now we calculate each term in Eq. (\ref{eq:dbmw11}). The first term is written as
\begin{eqnarray}
\delta w_{(11, \beta)}^{(1)} &=& v_{\alpha}H_{\alpha \beta}^{(1)}
 \;,
\end{eqnarray}
where the repeated indices imply the summation and
\begin{eqnarray}
H_{11}^{(1)}&=&-4\pi R_0 \sum_{n} c_n (t) i^{-n} \int_{\bm{q}}\frac{q_x^2}{q}G_q^2e^{in \theta}J_1(qR_0)J_n (q R_0) \nonumber \\
&=&a_{1}(c_2+c_{-2})
 \;,
\label{eq:H111}
\end{eqnarray}
with
\begin{eqnarray}
a_{1}=\frac{R_0}{2}\int_{0}^{\infty}dq q^2G_q^2 J_1(qR_0)J_2 (q R_0) 
 \;.
\label{eq:A1}
\end{eqnarray}
Similarly one obtains
\begin{eqnarray}
H_{12}^{(1)}&=&-4\pi R_0 \sum_{n} c_n (t)i^{-n} \nonumber \\
&\times&\int_{\bm{q}}\frac{q_xq_y}{q}G_q^2e^{in \theta} J_{1}(qR_0)J_n (q R_0) \nonumber \\
&=&a_1(c_2-c_{-2})i
 \;.
\label{eq:H122}
\end{eqnarray}
together with the relations $H_{22}^{(1)}=-H_{11}^{(1)}$ and $H_{12}^{(1)}=H_{21}^{(1)}$. It  should be noted that because of the factors $q_x^2$ in (\ref{eq:H111})  and $q_xq_y$ in  (\ref{eq:H122}), only the components $n=\pm 2$ contribute to $H_{11}^{(1)}$ and $H_{12}^{(1)}$.

The second term in Eq. (\ref{eq:dbmw11}) can be written by using the formula (\ref{eq:form}) as
\begin{eqnarray}
\delta w_{(11, \beta)}^{(2)} &=& v_{\alpha}H_{\alpha \beta}^{(2)}
 \;,
\end{eqnarray}
where
\begin{eqnarray}
H_{\alpha \beta}^{(2)}&=&\frac{2}{R_0} \sum_{n} c_n (t) i^{n} \int_{\bm{q}}q_{\alpha}G_q^2\theta_q^{(0)}\frac{\partial}{\partial q_{\beta}}e^{in \theta} \nonumber \\
&\times&\big[qJ_{n-1}(qR_0)+\frac{1-n}{R_0}J_n (q R_0)\big] \nonumber \\
&=&-\frac{2}{R_0} \sum_{n} c_n (t) i^{n} \int_{\bm{q}}e^{in \theta}\big[qJ_{n-1}(qR_0) \nonumber \\
&+&\frac{1-n}{R_0}J_n (q R_0)\big] \frac{q_{\alpha}q_{\beta}}{q}\frac{\partial}{\partial q}(G_q^2\theta_q^{(0)})
 \;.
\label{eq:Hab2}
\end{eqnarray}
In the derivation of the second line, we have used the fact $c_0=0$ which comes from the domain area conservation. Each component of Eq. (\ref{eq:Hab2}) is readily obtained as 
\begin{eqnarray}
H_{11}^{(2)}&=&a_2(c_2+c_{-2})  \;, \nonumber \\
H_{12}^{(2)}&=&a_2(c_2-c_{-2})i
 \;,
\label{eq:H112}
\end{eqnarray}
where $H_{22}^{(2)}=-H_{11}^{(2)}$ and $H_{12}^{(2)}=H_{21}^{(2)}$  and
\begin{eqnarray}
a_{2}&=&\frac{1}{4\pi R_0}\int_{0}^{\infty}dq q^2\big[qJ_1(qR_0) \nonumber \\
&-&\frac{1}{R_0}J_2 (q R_0)\big] \frac{\partial}{\partial q}(G_q^2\theta_q^{(0)})
 \;.
\label{eq:A2}
\end{eqnarray}

\begin{figure}[hbpt]
\begin{center}
\includegraphics[width=0.6\linewidth]{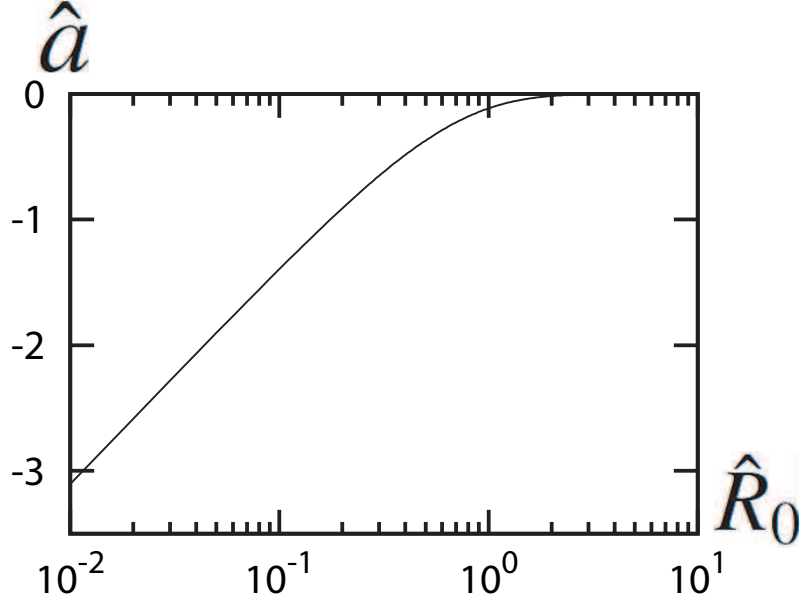}
\caption{The coefficient $\hat{a}=aD^2/R_0^2$ as a function of $\hat{R}_0=\kappa R_0$.}
\label{a}
\end{center}
\end{figure}

Putting the expressions of $H_{\alpha \beta}^{(n)}$ with $n=1$ and 2  together, Eq.  (\ref{eq:vvv3}) is finally written as
\begin{equation}
m\dot{v}_{\alpha} +\frac{1}{2}(\tau-\tau_c)v_{\alpha}  +gv_{\alpha}|\bm{v}|^2=-av_{\beta}S _{\beta \alpha}
 \;,
\label{eq:vvv4}
\end{equation}
where 
\begin{eqnarray}
a=2(a_1+a_2)
 \;,
\label{eq:a}
\end{eqnarray}
and the tensor $S _{\beta \alpha}$ has been defined in Eq. (\ref{eq:tenS}) with  (\ref{eq:S}).
The coefficient $a$ is evaluated numerically and is displayed as  a function of $\hat{R}_0=R_0\kappa$ in Fig. \ref{a}. It is noted that the coefficient $a$ is negative for $0 < \hat{R}_0 <\infty$. When $\hat{R}_0 \ll1$, one has an analytical expression $\hat{a}=aD^2/R_0^2=-(3/4) |\ln \hat{R}_0|$. The numerical result in Fig. \ref{a} is consistent with this.

%%%%%%%%%%%%%%%%%%%%%%%%%%%%%%%%%%%%%%%%%%
\section{Equations of motion for $c_{\pm2}(t)$ and $c_{\pm3}(t)$ }

In order to derive the time-evolution equations for $c_{\pm2}(t)$ and $c_{\pm3}(t)$, we start with the interface equation of motion (\ref{eq:interface}). Since it will be shown in section 7  that $c_{\pm2}(t)\sim O(\bm{v}^2)$  and $c_{\pm3}(t)\sim O(\bm{v}^3)$, the coupling such as $c_nc_m$ is negligible for slow dynamics of a domain. Therefore, we may consider only the linear order of the deviation $\delta R$. (This takes account of the coupling between $c_n$ and $\bm{\rho}$.) Linearizing Eq. (\ref{eq:interface}) with respect to the deformation, one obtains
\begin{eqnarray}
\tau\frac{\partial \delta R}{\partial t}=\frac{\epsilon}{R_0^2}\big(\frac{\partial ^2 \delta R}{\partial \phi^2}+\delta R \big) -4\delta w +L
 \;,
\label{eq:linear}
\end{eqnarray}
where $\delta w=w-w_0$ with $w_0$ the value of $w$ for a circular domain and we have used the fact that $\tau dc(w_0)/dw_0=-4$ as shown from Eq. (\ref{eq:velocity}).  The last term $L$ is necessary in order  to eliminate any zero modes which might arise from $\delta w$.

\begin{figure}[hbpt]
\begin{center}
\includegraphics[width=0.6\linewidth]{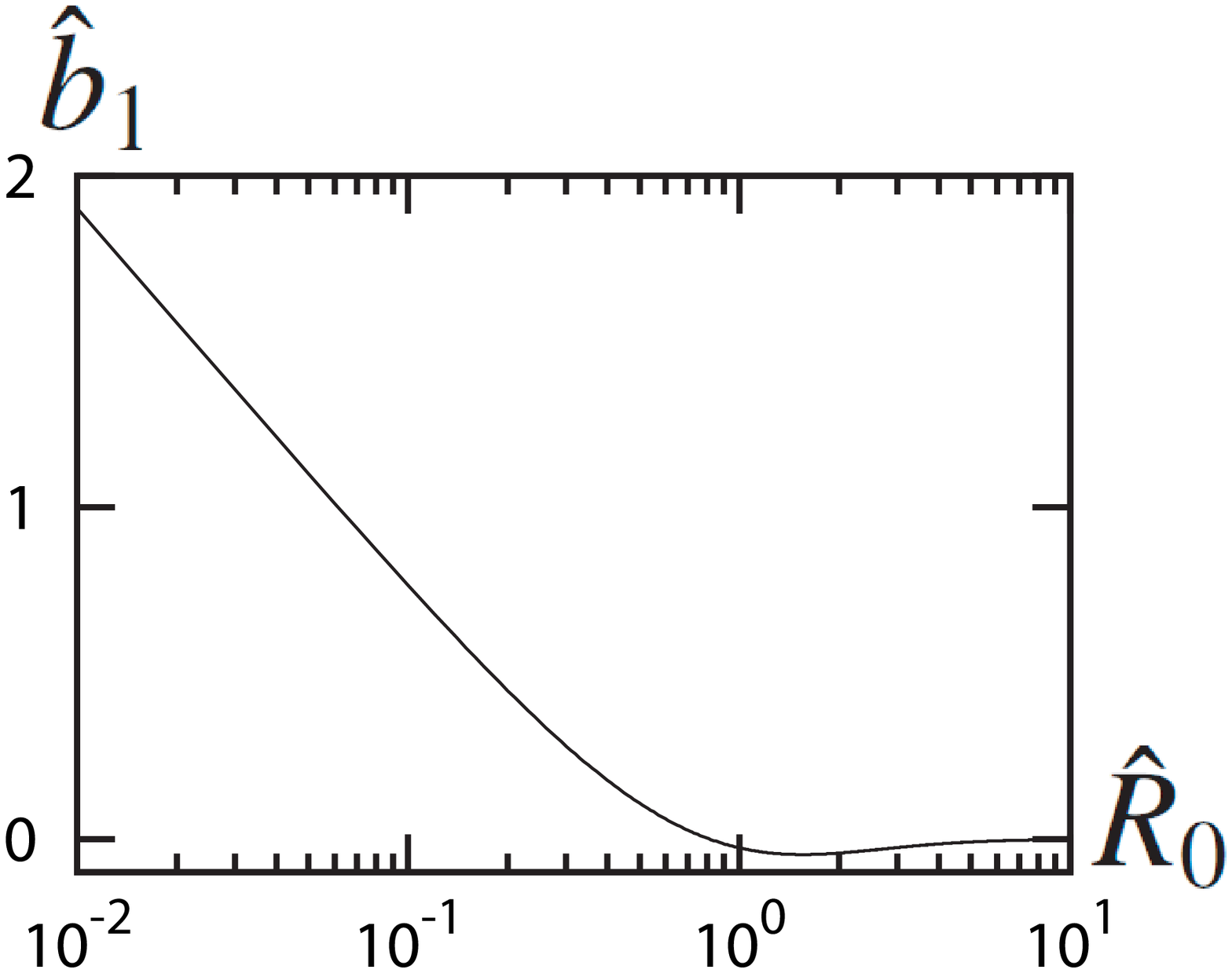}
\caption{The coefficient $\hat{b}_1=D^2b_1 /R^2_0$ as a function of $\hat{R}_0=\kappa R_0$.}
\label{b1}
\end{center}
\end{figure}

First, we consider the $n=\pm2$ modes. From Eqs. (\ref{eq:w0q}), (\ref{eq:w1q}) and (\ref{eq:w2q}), one notes that 
there are three terms in $\delta w$, which produce $n=\pm 2$ modes;
\begin{eqnarray}
\delta w=\delta w_0+\delta w_1+\delta w_2
%+\delta w_3
 \;,
\label{eq:dw}
\end{eqnarray}
where
\begin{eqnarray}
\delta w_0&=&\int_{\bm{q}}G_q \theta_qe^{i\bm{q}\cdot\bm{R}(\omega)}-\int_{\bm{q}}G_q \theta_q^{(0)}e^{i\bm{q}\cdot\bm{R}^{(0)}(\omega)}\label{eq:dw0}  \;, \\
\delta w_1&=&\delta w_{11}+\delta w_{12} \nonumber \\
&=&i \int_{\bm{q}} (\bm{q} \cdot \bm{v}) G_{q}^{2} \theta_q e^{i \bm{q}\cdot \bm{R}(\omega)}
-\int_{\bm{q}}G_q^2 \frac{\partial \theta_q}{\partial t}e^{i\bm{q}\cdot\bm{R}(\omega)}
 \;,
\label{eq:dw1} \\
%\delta w_2&=&\int_{\bm{q}}G_q^3 \frac{\partial^2 \theta_q}{\partial t^2}e^{i\bm{q}\cdot\bm{R}(\omega)}
% \;,
%\label{eq:dw2} \\
\delta w_2&=& - \int_{\bm{q}} (\bm{q} \cdot \bm{v})^2 G_{q}^{3} \theta_q e^{i \bm{q}\cdot \bm{R}(\omega)} 
 \;.
\label{eq:dw3}
\end{eqnarray}
Here we have ignored the term with $\partial^2 \theta_q/\partial t^2$ which causes $\ddot{c}_n$ and are higher order in the final equation of motion.
By using Eq. (\ref{eq:theta1}), these are readily evaluated as
\begin{eqnarray}
\delta w_0&=&-\sum_nD_nc_n e^{in\phi}\label{eq:dw01}  \;, \\
\delta w_{11}&=&-\frac{1}{2}(v_1-iv_2)\sum_nA_nc_n e^{i(n+1)\phi} \nonumber \\
&-&\frac{1}{2}(v_1+iv_2)\sum_nB_nc_n e^{i(n-1)\phi} 
\label{eq:dw11}  \;, \\
\delta w_{12}&=&-\sum_nE_n\dot{c}_n e^{in\phi} \label{eq:dw11}  \;, \\
%\delta w_2&=&\sum_nF_n\ddot{c}_n e^{in\phi} \label{eq:dw21}   \;, \\
\delta w_2&=& -\frac{G_0}{2}\bm{v}^2 +\frac{G_1}{2}\big[\big(\frac{1}{2}(v_1^2-v_2^2)-iv_1v_2\big)e^{2i\phi}   \nonumber \\
&+&\big(\frac{1}{2}(v_1^2-v_2^2)+iv_1v_2\big)e^{-2i\phi}\big]  \;,
\label{eq:dw31}
\end{eqnarray}
where 
\begin{eqnarray}
A_n&=&R_0\int_0^{\infty}dq q^2 G_q^2[J_n(qR_0)J_{n+1}(qR_0) \nonumber \\
&+&J_1(qR_0)\frac{\partial}{\partial qR_0}J_1(qR_0)]\label{eq:An}  \;,  \\
B_n&=&R_0\int_0^{\infty}dq q^2 G_q^2[-J_n(qR_0)J_{n-1}(qR_0) \nonumber \\
&+&J_1(qR_0)\frac{\partial}{\partial qR_0}J_1(qR_0)]\label{eq:Bn}  \;,  \\
D_n&=&R_0\int_0^{\infty}dq q G_q[J_1(qR_0)^2-J_n(qR_0)^2]\label{eq:Dn}  \;,  \\
E_n&=&R_0\int_0^{\infty}dq q G_q^2J_n(qR_0)^2\label{eq:En}  \;, \\
%F_n&=&R_0\int_0^{\infty}dq q G_q^3J_n(qR_0)^2\label{eq:Fn}  \;, \\
G_{\ell}&=&R_0\int_0^{\infty} dq q^2 G_{q}^{3} J_{\ell}(qR_0) J_{\ell+1}(qR_0)  \;.
\label{eq:Gell}
\end{eqnarray}
Note  the relation that $A_{-n}=B_n$.
It is also noted from Eq. (\ref{eq:dw31}) that $\delta  w_2$ contains the $n=0$ mode proportional to $\bm{v}^2$, which should be absorbed into the Lagrange multiplier $L$ in Eq. (\ref{eq:linear}). From the $n=\pm 2$ modes, one obtains
\begin{eqnarray}
\delta w_2^{n=2}+\delta w_2^{n=-2}&=&\frac{G_1}{2}(v_1^2-v_2^2)\label{eq:deltaweR}  \;, \\
i(\delta w_2^{n=2}-\delta w_2^{n=-2})&=&G_1v_1v_2  \;.
\label{eq:deltaweI}  
\end{eqnarray}
After extracting the $n=\pm 2$ terms from Eq. (\ref{eq:linear}),  equation for the tensor $S$ defined by Eq.   (\ref{eq:tenS}) is given by
\begin{eqnarray}
%m_2\frac{d^2 S_{\alpha \beta}}{d t^2}+
\Gamma_2\frac{d S_{\alpha \beta}}{d t}=-K_2S_{\alpha \beta}+b[v_{\alpha}v_{\beta}-\frac{\delta_{\alpha \beta}}{2}\bm{v}^2] + b_1 U_{\alpha \beta \gamma}v_{\gamma} \;,
\label{eq:eqc2}
\end{eqnarray}
where 
%$m_2=4F_2$, 
$\Gamma_2=\tau-4E_2$, $b=-4G_1$, 
%\begin{eqnarray}
$K_2=(3\epsilon/R_0^2)-4D_2 $ and $b_1=2B_3$.
%上の？？？を記入
%\label{eq:K2}
%\end{eqnarray}
It is noted that the 
coefficient 
%$m_2$ is positive whereas 
$b$ is negative. In the vicinity of the bifurcation $\tau \sim \tau_c$ with $\tau_c$ given by (\ref{eq:tauc}), the coefficient $\Gamma_2$ is positive.  The coefficient $K_2$ becomes negative for sufficiently large values of $R_0$ indicating an instability of a motionless circular domain \cite{Ohta89}. The coefficient  $b_1$ is plotted as a function of $\hat{R}_0$ in Fig. \ref{b1}. It is positive for $\hat{R}_0 < 0.8$  and negative for $\hat{R}_0 >0.8$.

\begin{figure}[hbpt]
\begin{center}
\includegraphics[width=0.6\linewidth]{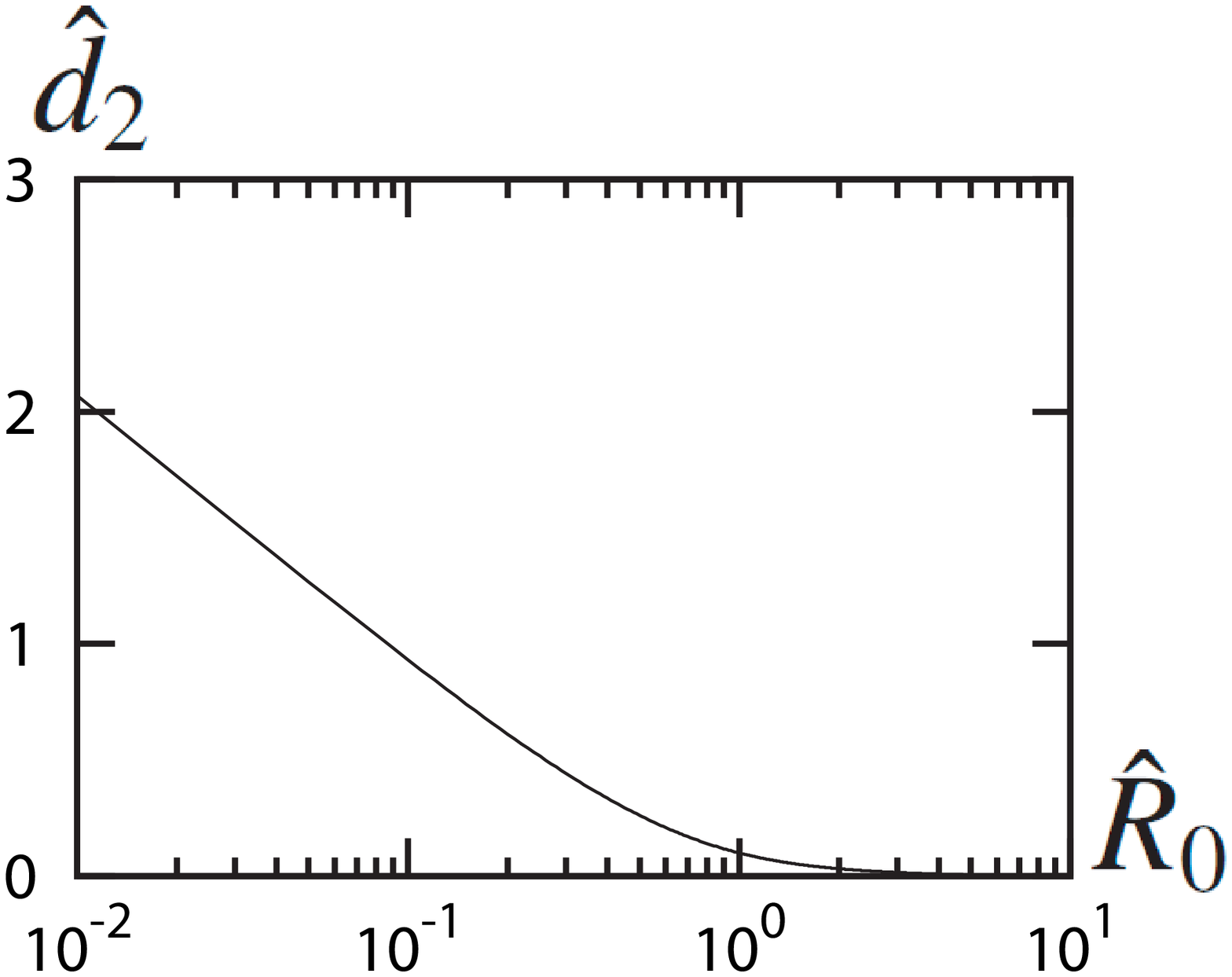}
\caption{The coefficient $\hat{d_2}=d_2D^2/R^2_0$ as a function of $\hat{R}_0=\kappa R_0$.}
\label{d2}
\end{center}
\end{figure}

By putting all the $n=\pm 3$ terms together,
the time-evolution equations for $c_{\pm3}$ or $T_i$ ($i=1$ and 2) defined by  (\ref{eq:T}) take the following form
\begin{eqnarray}
%m_3\frac{d^2 T_{1}}{d t^2}+
\Gamma_3\frac{d T_1}{d t}&=&-K_3T_1 \nonumber \\
&+&d_1v_{1}(v_{1}^2-3v_{2}^2)+d_2(v_{1}S_{11}-v_{2}S_{12})  \;,
\label{eq:T1} \\
%m_3\frac{d^2 T_{2}}{d t^2}+
\Gamma_3\frac{d T_2}{d t}&=&-K_3T_2 \nonumber \\
&-&d_1v_{2}(v_{2}^2-3v_{1}^2)-d_2(v_{2}S_{22}-v_{1}S_{21})  \;,
\label{eq:T2} 
\end{eqnarray}
where
% $m_3=4F_3$, 
$\Gamma_3=\tau-4E_3$ and $K_3=8\epsilon/R_0^2-4D_3$ and the relation $S_{11}=-S_{22}$ has been used. The second term in 
Eqs.  (\ref{eq:T1}) and  (\ref{eq:T2}) arises from  $w_{(3)}$ given by (\ref{eq:w3q}) whereas the third term in Eqs.  (\ref{eq:T1}) and  (\ref{eq:T2}) comes from $w_{(11)}$ given by (\ref{eq:w1q}) and the coefficients $d_1$ and $d_2$ are given, respectively, by
\begin{eqnarray}
d_1&=&R_0\int_{0}^{\infty}dqq^3G_q^4J_1(qR_0)J_3(qR_0)   \;,
\label{eq:d1} \\
d_2&=&2R_0\int_{0}^{\infty}dqq^2G_q^2\big[J_2(qR_0)J_3(qR_0) \nonumber \\
&+&J_1(qR_0)\frac{\partial}{\partial qR_0}J_1(qR_0)\big]  \;.
\label{eq:d-2} 
\end{eqnarray}
The coefficient $d_2$ is displayed as a function of $\hat{R_0}$ in Fig. \ref{d2}. Both $d_1$ and $d_2$ are positive.

Equations (\ref{eq:T1}) and (\ref{eq:T2}) can be represented in terms of the third-rank tensor $U_{\alpha\beta\gamma}$ defined by (\ref{eq:tenU}) as
\begin{eqnarray}
%&m_3&\frac{d^2 U_{\alpha\beta\gamma}}{d t^2}+
&\Gamma_3&\frac{d U_{\alpha\beta\gamma}}{d t}=-K_3U_{\alpha\beta\gamma} \nonumber \\
&+&4d_1\big[v_{\alpha}v_{\beta}v_{\gamma}-\frac{v_{\eta}v_{\eta}}{4}(\delta_{\alpha\beta}v_{\gamma}+\delta_{\beta\gamma}v_{\alpha}+\delta_{\gamma\alpha}v_{\beta})\big] \nonumber \\
&+&\frac{2d_2}{3}\big[S_{\alpha\beta}v_{\gamma}+S_{\beta\gamma}v_{\alpha}+S_{\gamma\alpha}v_{\beta} \nonumber \\
&-&\frac{v_{\eta}}{2}(\delta_{\alpha\beta}S_{\gamma\eta}+\delta_{\beta\gamma}S_{\alpha\eta}+\delta_{\gamma\alpha}S_{\beta\eta})\big]  \;.
\label{eq:EqU}
\end{eqnarray}
Other terms such as $U\bm{v}\bm{v}$ are higher order as shown in the next section.

%A more general expression including other possible couplings in arbitrary dimensions is given in Appendix.

%%%%%%%%%%%%%%%%%%%%%%%%%%%%%%%%%%%%%%%%%%%%%%%%%%%
\section{Order estimation and stationary solutions}

In this section, we shall discuss how to justify the approximations employed in the derivations of the time-evolution equations Eqs. (\ref{eq:vvv4}),  (\ref{eq:eqc2}) and (\ref{eq:EqU}). As emphasized, the basic assumption is that we are concerned with the domain dynamics near the supercritical drift bifurcation $\tau=\tau_c$. Therefore, the smallness parameter is $\delta\equiv \tau_c-\tau$. It is readily shown from Eq. (\ref{eq:vvv4}) that all the terms are of the order of $\delta^{3/2}$ since $\bm{v}\sim O(\delta^{1/2})$ and time is scaled as $\hat{t}= t\delta$ and  $S\sim O(\delta)$ as can be seen from the first and second terms in  Eq. (\ref{eq:eqc2}). Similarly, one notes from Eq. (\ref{eq:EqU}) that $U\sim O(\delta^{3/2})$. 

The above order estimation tells us that the terms ignored in Eq. (\ref{eq:vvv4}) are indeed higher order in $\delta$. For example, $d^2\bm{v}/dt^2\sim O(\delta^{5/2})$ and $SS\bm{v}\sim O(\delta^{5/2})$. On the other hand, the first and second terms in Eq. (\ref{eq:eqc2}) are of the order of $O(\delta)$ whereas $dS/dt\sim O(\delta^2)$ and $U\bm{v}\sim  O(\delta^2)$ and all other terms ignored are higher order. Equation (\ref{eq:EqU}) has also a similar property. Therefore, up to the leading order, the time derivative in $S$ and $U$ should be ignored. This is not surprising since, for $\delta \to 0$, the center of gravity is a slow variable but the deformations around a circular shape are not necessarily  slow  and might be eliminated adiabatically. This fact does not cause any difficulties in the study of the stationary shape of a steadily propagation domain.

It is more convenient to employ the following representation
\begin{eqnarray}
v_1 &=& v \cos \phi, \ \ v_2 = v \sin \phi  \;, \\
S_{11} &=& \frac{1}{2} s \cos 2 \theta  \;,\ \ S_{12} = \frac{1}{2} s \sin 2 \theta  \;, \\
U_{111} &=& c_{3} + c_{-3}, \ \ U_{222} = -i(c_{3} - c_{-3})  \;, \\
c_{3} &=& \frac{\delta_3}{2} e^{-3i \phi_3} = \frac{z}{4} e^{-3i\varphi}  \;,
\end{eqnarray}
where $v$ should not be confused with the inhibitor $v$ in Eq. (\ref{eq:peqv}).
Substituting these into
Eqs. (\ref{eq:vvv4}),  (\ref{eq:eqc2}) and (\ref{eq:EqU}) yields
\begin{eqnarray}
\frac{d}{dt} v&=& v(\gamma - v^{2}) - \frac{1}{2} a'sv \cos 2(\theta - \phi)  \;, \label{eq:v}\\
\frac{d}{dt} \phi &=& - \frac{1}{2} a's \sin 2 (\theta - \phi)  \;, \label{eq:phi}
\end{eqnarray}
\begin{eqnarray}
\frac{d}{dt} s &=& - \kappa s + b' v^{2} \cos 2 (\theta -\phi)  \;, \label{eq:s}\\
\frac{d}{dt} \theta &=& - \frac{b'v^{2}}{2s} \sin 2 (\theta -\phi)  \;, \label{eq:theta}
\end{eqnarray}
\begin{eqnarray}
\frac{d}{dt} z &=& -Kz + d'_1 v^{3} \cos 3 (\varphi - \phi) \nonumber \\
&+ & d'_2 sv \cos (3\varphi - 2\theta -\phi )  \;, \label{z}  \\
\frac{d}{dt} \varphi &=& - \frac{d'_1}{3z} v^{3} \sin 3 (\varphi - \phi) \nonumber \\
&-&\frac{d'_2}{3z} sv \sin (3 \varphi - 2\theta - \phi)  \;, \label{varphi}
\end{eqnarray}
where $\gamma=(\tau_c-\tau)/(2m)$, $a'=a/m$,  $\kappa=K_2/\Gamma_2$, $b'=b/\Gamma_2$, $K=K_3 / \Gamma_3$, $d'_1 = 2d_1/\Gamma_3$, and $d'_2 = d_2 / \Gamma_3$ and we have put $g/m=1$ without loss of generality.  We have ignored the $b_1$ term in Eq.  (\ref{eq:eqc2}) since it is higher order.  In what follows, we drop the prime in $a'$, $b'$, $d'_1$ and $d'_2$. 

It is noted that Eqs.  (\ref{eq:v})-(\ref{eq:theta}) are closed. In the previous paper, we have shown that there are two stable stationary solutions depending on the parameters \cite{Ohta09}. For simplicity, we restrict ourselves to $ab>0$. Here we consider a straight motion propagating, without loss of generality, along the $x$ axis, i.e., $\phi=0$. When $b$ is positive, $\theta=0$, and the steady value of the amplitudes is given by $v_0^2=\gamma/(1+B)$ and $s_0=bv_0^2/\kappa$ where $B\equiv ab/(2\kappa)$. whereas, when $b$ is negative, the solution is given by $\theta=\pi/2$, $v_0^2=\gamma/(1+B)$ and $s_0=-bv_0^2/\kappa$. 

As shown in the preceding section, the both constants $a$ and $b$ are negative in the reaction diffusion equations studied in this paper.  Therefore,  the long axis of an elliptical domain is perpendicular to the velocity vector.
Substituting the stationary solutions given above into  Eqs. (\ref{z}) and ({\ref{varphi}), one obtains  the time-independent solution
\begin{eqnarray}
Kz_0 &=& d_1 v_{0}^{3} \cos 3 \varphi_0 + d_2 s_0 v_0 \cos (3 \varphi_0 -\pi)  \;, \\
0 &=& d_1 v_{0}^{3} \sin 3 \varphi_0 + d_2 s_0 v_0 \sin (3 \varphi_0 -\pi)  \;.
\end{eqnarray}
%where $v_0 = \sqrt{\gamma /(1+B)}$ and $s_0 = - bv_{0}^{2}/\kappa$.
Since $z_0$ should be positive, we find the following condition irrespective of the sign of $b$;
\begin{eqnarray}
d_1 + \frac{b}{\kappa} d_2 > 0 \ \Rightarrow \ \cos 3 \varphi_0 = 0 \ \Rightarrow \ \varphi_0 = 0  \;,  \label{eq:posi}\\
d_1 + \frac{b}{\kappa} d_2 <0 \ \Rightarrow \  \cos 3\varphi_0 = -1 \ \Rightarrow \ \varphi_0 = \pi  \;, \label{eq:nega}
\end{eqnarray}
and
\begin{equation}
z_0 = \left| d_1 + \frac{b}{\kappa} d_2 \right| \frac{v_{0}^{3}}{K}  \;.
\end{equation}

The stationary shape of a domain propagating along the $x$ axis ($\phi=0$) is shown in Fig. \ref{shape}(a) for $\varphi_0=0$. It should be noted that in the present reaction diffusion system, the constants $d_1$ and $d_2$ are positive whereas $b$ is negative. When  $\kappa$ is large, the straight motion is stable and the inequality (\ref{eq:posi}) always holds. Therefore, we have the deformed domain shown in Fig.  \ref{shape}(a). We emphasize that this is consistent with the numerical simulations in ref.  \cite{Mikhailov}. On the other hand, 
when $b$ is positive (the constant $a$ is also assumed to be positive), the long axis of an elliptical domain is parallel to the velocity vector \cite{Ohta09}. In this case, the stationary domain shape is displayed in Fig. \ref{shape}(b) for $\varphi_0=0$. If the condition (\ref{eq:nega}) holds, i.e., $\varphi_0=\pi$, the shape of domain is given by the mirror symmetry with respect to the vertical axis of those in Fig. \ref{shape}.

\begin{figure}[hbpt]
\begin{center}
\includegraphics[width=.6\linewidth]{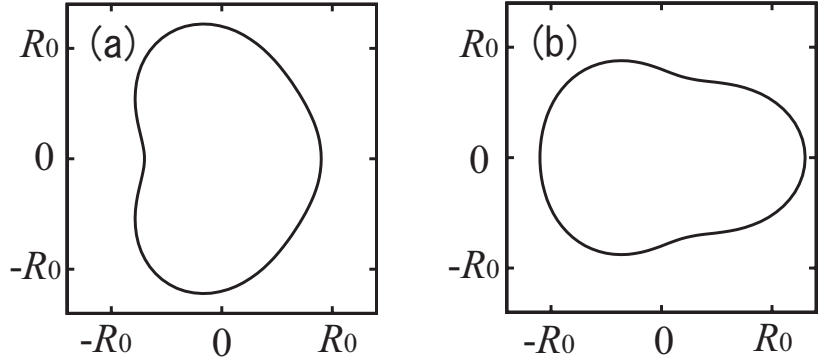}
\caption{(a) The stationary shape of a domain traveling along the $x$ axis ($\phi=0$) for $\varphi_0 = 0$ and (a) $\theta = \pi/2$, and (b) $\theta = 0$.}
\label{shape}
\end{center}
\end{figure}

Hereafter, we shall discuss the properties of the time-evolution equations  (\ref{eq:vvv4}),  (\ref{eq:eqc2}) and (\ref{eq:EqU}) from more general point of view.  In fact, all the terms are derived by considering the possible couplings which can be generated from the first rank tensor (vector) $ \bm{v}$, the second rank tensor $S$ and the  third rank tensor $U$ not necessarily restricted to the reaction-diffusion equations. 
%See also Appendix. 
Therefore, if the coefficient $\Gamma_2 \sim O(\delta^{-1})$ (and $\Gamma_3 \sim O(1)$), one has to retain the $dS/dt$ term in $\Gamma_2$ whereas the $U\bm{v}$ term of the order of $O(\delta^2)$ can be ignored. The resulting coupled equations  (\ref{eq:vvv4}) and  (\ref{eq:eqc2}) are the same as those introduced in ref. \cite{Ohta09} where the rotation bifurcation has been found. The bifurcation occurs at 
\begin{eqnarray}
\gamma=\gamma_c= \frac{\kappa^2 }{ab}+\frac{\kappa}{2}  \;,
\label{eq:gammac}
\end{eqnarray}
such that the stationary straight motion becomes unstable for $\gamma \ge \gamma_c$. 
Assuming a circular motion appears for $\gamma \ge \gamma_c$, we put $v=v_r$, $s=s_r$, 
$
\theta = \omega t+\zeta/2
$ and $ \phi = \omega t$.
Substituting these into Eqs.    (\ref{eq:v})-(\ref{eq:theta}),  one obtains
after some algebra 
$
v_r^2 =\gamma-\kappa/2
$
,
$s_r^2 = bv_r^2/a$,
%\label{sr}\\
$\cos \zeta =\kappa/as_r$
%\label{ss}\\
and 
$\omega^2 =(ab/4)(v_r^2-v_c^2)$ where $v_c=\kappa/(ab)^{1/2}$.
%\label{omegar}
%\end{eqnarray}
It has been shown that the frequency $\omega$ continuously increases from zero at $\gamma =\gamma_c$. 
We have carried out the stability of the straight motion analytically and  that of the circular motion numerically and have found that when the straight motion is stable, the circular motion is unstable and vise versa. That is, there is no parameter regime where these two motions coexist \cite{Ohta09}.

The stationary shape of a circular motion can also be obtained without any difficulty.
There are two types of circular motion; clockwise and anti-clockwise rotation. We consider only the anti-clockwise rotation putting 
%\begin{eqnarray}
%\theta &=& \omega t + \zeta /2, \\
$\varphi = \omega t + \delta /3$.
%\phi &=& \omega t.
%\end{eqnarray}
Then, the set of equations (\ref{z}) and ({\ref{varphi}) become
\begin{eqnarray}
Kz_r &=& d_1 v_{r}^{3} \cos \delta + d_2 s_r v_r \cos (\delta - \zeta)  \;, \\
3z_r \omega &=& -d_1 v_{r}^{3} \sin \delta - d_2 s_r v_r \sin (\delta - \zeta)  \;.
\end{eqnarray}
From these two equations, the stationary solution is given by
\begin{eqnarray}
\frac{\cos \delta}{ z_{r}}&=& \frac{\left(d_1 v_{r}^{2} + \kappa d_3\right)K - 6 \omega^{2}d_3}{d_{1}^{2} v_{r}^{5} + d_3 v_{r}^{3}\left(2 \kappa d_1 + b d_2 \right)}  \;,  \label{eq:zcos}\\
\frac{\sin \delta}{ z_{r}}&=& - \frac{2\omega d_3 K + 3 \omega \left(d_1 v_{r}^{2} + \kappa  d_3\right)}{d_{1}^{2} v_{r}^{5} +d_3 v_{r}^{3} \left(2 \kappa d_1 + b d_2\right)}  \;,  \label{eq:zsin}
\end{eqnarray}
where $d_3=d_2/a$ and 
\begin{equation}
\tan \delta = - \frac{2\omega d_3 K + 3 \omega \left(d_1 v_{r}^{2} + \kappa d_3\right)}{\left(d_1 v_{r}^{2} + \kappa d_3\right)K - 6 \omega^{2} d_3}  \;.
\end{equation}
%where we have used the facts that $s_r \sin \zeta = -2 \omega/a$, $s_r \cos \zeta = \kappa/a$, and $4 \omega^{2} + \kappa^{2} = ab v_{r}^{2}$. 
From Eqs. (\ref{eq:zcos}) and (\ref{eq:zsin}), one has
\begin{equation}
z_{r}^{2} = \frac{d_{1}^{2} v_{r}^{6} + d_3 v_{r}^{4} \left(2 \kappa d_{1} + b d_2\right)}{K^{2} + 9 \omega^{2}}.
\end{equation}
The replacements  $\omega \to - \omega$, $\zeta \to - \zeta$, and $\delta \to -\delta$ give us the solution for a clockwise rotation.

%\begin{figure}[hbpt]
%\begin{center}
%\includegraphics[width=0.8\linewidth]{fig2.eps}
%\caption{(a) Stability diagram on the $\gamma-\kappa$ plane for $ab=0.5$. (b) The velocity $v$ (full line) and the frequency $\omega$ (broken line) as a function of $\gamma$ for $a=-1.0,b=-0.5$ and $\kappa=0.5$. The bifurcation occurs at $\gamma_c = 0.75$.}
%\label{critical}
%\end{center}
%\end{figure}

\section{Summary and discussion}

In this paper, we have studied the domain dynamics in an excitable reaction-diffusion system with the global coupling. The set of time-evolution equations has been derived in the singular limit that the interface width is infinitesimal. The main results are Eqs. (\ref{eq:vvv4}), (\ref{eq:eqc2}) and (\ref{eq:EqU}). The coefficient $a$ in Eq. (\ref{eq:vvv4}) and $b$ in Eq.  (\ref{eq:eqc2}) are negative whereas the coefficients $d_1$ and $d_2$ in Eq.  (\ref{eq:EqU}) are positive. Therefore, the deformed shape in the steady state takes the form shown  in Fig. \ref{shape}(a)). We emphasize that this result is consistent with the computer simulations of  Eqs. (\ref{eq:pequ}) and (\ref{eq:peqv}) with (\ref{eq:peqf}) and
(\ref{eq:global})  \cite{Mikhailov}. 

The present theory is restricted to two dimensions. Extension to three dimensions is possible if one expands  deformations around a spherical shape in terms of spherical harmonics. Representation of the set of equations by introducing some tensor variables as in the present theory is also convenient. However, it is expected that a propagating domain in three dimensions are elongated not only to a rod shape or a cone shape (corresponding to the form in Fig. \ref{shape}(b)) parallel to the propagating direction but also to a jellyfish-like shape (corresponding to the form in Fig. \ref{shape}(a)) when the elongation is perpendicular to the propagating direction.  Therefore, the direct use of the same tensors $S_{\alpha \beta}$ for the nematic order parameter and $U_{\alpha \beta \gamma}$ for banana-shape liquid crystals seems not to be possible.

%Here we discuss the conditions that the approximations employed are justified . In the stationary situation, the right hand side of Eq.  (\ref{eq:eqc2})  indicates that $S\sim O(\bm{v}^2)$ and that of Eq.(\ref{eq:EqU}) implies $U\sim O(\bm{v}^3)$. Therefore, one notes that all the couplings up to $O(\bm{v}^3)$ are consistently taken into account in Eqs.  (\ref{eq:vvv4}), (\ref{eq:eqc2}) and (\ref{eq:EqU}). In the time-dependent situation, there must be terms like $\dot{S}\bm{v}$ in Eq. (\ref{eq:EqU}).  However, if the system is close to the rotation bifurcation, the acceleration is small enough so that the higher order time derivatives can be ignored. In this sense, the inertia term $m\bm{\dot{v}}$ in Eq.  (\ref{eq:vvv4}) should be retained whereas the first term in Eq.(100) and (106) may be ignored.
% It should be noted that the tensor $U$ does not enter in Eqs. (\ref{eq:eqc2}) and (\ref{eq:EqU}) up to the leading order. Therefore the previous analysis based only on $\bm{v}$ and $S$ is unaltered by introducing $U$. However, this third rank tensor is required to express the asymmetry between the head and the rear of a propagating domain as shown in section 7.
 
Apart from the generalization to three dimensions, there are several important open problems. First of all, 
 %the rotation bifurcation should be verified by numerical simulations of the original reaction-diffusion equations (\ref{eq:pequ}) and (\ref{eq:peqv}) with  (\ref{eq:peqf}) and (\ref{eq:global}). Since the bifurcation threshold is obtained as Eq. (\ref{eq:gammac}), numerical search of circular motion of a domain would not be so difficult. Secondly, 
 the present theory should be generalized to a collision process  of a pair of deformable self-propelled domains.
 If the interaction is repulsive, the motion of the two domains becomes slow as approaching to each other and hence those shapes are expected to change as shown in the present theory. However, the existence of another domain itself tends to deform generally the domain, which should be incorporated into our theory. Next, the dynamics of domains undergoing the circular motion will be investigated in detail when the global orientational coupling is imposed. Since a clock-wise motion and a counter clock-wise motion coexists, the orientational ordering is interrupted so that complex dynamics, such as synchronization, localization and chaotic motions appear depending on the parameters \cite{Ohta09}. Numerical simulations and a theoretical study reducing to non-linear coupled oscillators are now being carried out \cite{Ohkuma09}. Third, the helical motion in three dimensions which is an extension of the circular motion in two dimensions must be investigated both numerically and analytically.  
 Finally, the stochastic dynamics of deformable self-propelled domains will be developed by adding random noise terms in the equations derived here. As a related study, quite recently, Sano et al. \cite{Sano09} have  introduced and investigated a dynamical model for the motion of amoeboid cell to analyze the experimental results obtained.
We hope to publish these investigations somewhere in the near future.

\subsection*{Acknowledgment}
This work was supported by 
the Grant-in-Aid for the priority area "Soft Matter Physics" 
and
the Global COE Program "The Next Generation of 
Physics, Spun from Universality and Emergence" both
from the Ministry of Education, Culture, Sports, Science and Technology (MEXT) of Japan.

%\clearpage

%\setlength{\baselineskip}{0mm}


\begin{thebibliography}{10}

\bibitem{Kapralbook}
R. Kapral, K. Showalter (eds.), 
\newblock  {\it Chemical Waves and Patterns} (Kluwer, Dordrecht, 1995)
\bibitem{Swinneybook}
 V. Krinsky, H. Swinney (eds.), 
\newblock  {\it Wave and Patterns in Biological and Chemical Excitable Media}
(North-Holland, Amsterdam, 1991)

\bibitem{PismenBook}
L. M. Pismen, 
\newblock  {\it Patterns and Interfaces in Dissipative Dynamics},
\newblock    (Springer, Berlin, 2006).

\bibitem{Swinney1}
 K. J. Lee, W. D. McCormick, J. E. Pearson and H. L. Swinney,
\newblock  Nature {\bf 369},  215 (1994).
\bibitem{Swinney2}
 K. J. Lee and H. L. Swinney,
\newblock  Phys. Rev. E{\bf 51}, 1899  (1995).

\bibitem{Pearson}
 J. E. Pearson, 
\newblock  Science {\bf 261}, 189  (1993).
\bibitem{Reynolds1}
 W. N. Reynolds, J. E. Pearson and S. Ponce-Dawson,
\newblock  Phys. Rev. Lett. {\bf 72},  2797 (1994).
\bibitem{Reynolds2}
 W. N. Reynolds, S. Ponce-Dawson and J. E. Pearson,
\newblock  Phys. Rev. E{\bf 56},  185 (1997).

\bibitem{Petrov}
 V. Petrov, S. K. Scott and K. Showalter, 
\newblock   Phil. Trans. R. Soc. Lond. A  {\bf 347}, 631 (1994).


\bibitem{Mikhailov}
 K. Krischer and A. Mikhailov, 
\newblock   Phys. Rev. Lett. {\bf 73},  3165 (1994).

\bibitem{Hayase}
 Y. Hayase, 
\newblock   J. Phys. Soc. Jpn. {\bf 66},  2584 (1997).
\bibitem{Hayase2}
Y. Hayase and T. Ohta, 
\newblock  Phys. Rev. Lett., {\bf 81}, 1726 (1998).
\bibitem{Hayase3}
Y. Hayase and T. Ohta, 
Phys. Rev. E {\bf 62}, 5998 (2000).

\bibitem{Ohta97}
 T. Ohta, J. Kiyose and M. Mimura, 
\newblock   J. Phys. Soc. Jpn. {\bf 66},  1551 (1997).
\bibitem{Ei}
 S.-I. Ei, M. Mimura and M. Nagayama, 
 Physica D  {\bf 165}, 176 (2002).
 
 \bibitem{Schenk}
 C. P. Schenk, M. Or-Guil, M. Bode and H.-G. Purwins, 
\newblock   Phys. Rev. Lett. {\bf 78}, 3781  (1997).
\bibitem{Or-Guil}
M. Or-Guil, M. Bode, C. P. Schenk, and H.-G. Purwins, Phys. Rev. E {\bf 57}, 6432 (1998).

\bibitem{Ohta01}
T. Ohta, Physica D  {\bf 151}, 61 (2001).

\bibitem{Nishiura05}
Y. Nishiura, T. Teramoto, and K. Ueda, Chaos  {\bf 15}, 047509 (2005).
\bibitem{Pismen01}
L. M. Pismen, Phys. Rev. Lett.  {\bf 86}, 548 (2001).

\bibitem{Schweitzer}
F. Schweitzer, W. Ebeling and B. Tilch, Phys. Rev. Lett.  {\bf 80}, 5044 (1998).
\bibitem{Condat}
C. A. Condat and G. J. Sibona, Physica D  {\bf 168}, 235 (2002).
\bibitem{Ebelingbook}
W. Ebeling and I. M. Sokolov, {\it Statistical Thermodynamics and Stochastic Theory of Nonequilibrium Systems} (World Scientific Publishing, London, 2005) 


\bibitem{Zykov08}
V. S. Zykov, Eur. Phys. J. Special Topics {\bf 157}, 209 (2008).

\bibitem{Ohta09}
T. Ohta  and T. Ohkuma, Phys. Rev. Lett.  {\bf 102}, 154101 (2009).

\bibitem{Prost}
P. G. de Gennes and J. Prost,  {\it The Physics of Liquid Crystals} (Oxford University Press, Oxford, 1993) 


\bibitem{Sumino}
Y. Sumino, N. Magome, T. Hamada, and K. Yoshikawa,  Phys. Rev. Lett.  {\bf 94}, 068301 (2005).
\bibitem{Nagai}
K. Nagai, Y. Sumino, H. Kitahata, and K. Yoshikawa,  Phys. Rev. E {\bf 71}, 065301(R) (2005).


\bibitem{Sano}
N. Shimoyama, K. Sugawara, T. Mizuguchi, Y. Hayakawa, and M Sano, Phys. Rev. Lett.  {\bf 76}, 3870 (1996).
\bibitem{Nagayama}
M. Nagayama, S. Nakata, Y. Doi, and Y. Hayashima,  Physica D  {\bf 194}, 151 (2004).
\bibitem{Baer}
K. John, M B\"ar, and U. Thiele, Eur. Phys. J. E  {\bf 18}, 183 (2005).
\bibitem{Golestanian}
R. Golestanian, T. B. Liverpool, and A. Ajdari, Phys. Rev. Lett.  {\bf 94}, 220801 (2005).
\bibitem{Pismen06}
L. M. Pismen, Phys. Rev. E {\bf 74}, 041605 (2006).
\bibitem{Kapral}
Y-G. Tao and R. Kapral, J. Chem. Phys. {\bf 128}, 164518 (2008).
\bibitem{Mikhailov05}
U. Erdmann, W. Ebeling and A. S. Mikhailov, Phys. Rev. E {\bf 71}, 051904 (2005).
\bibitem{Loewen}
S. van Teeffelen and H. L\"owen, Phys. Rev. E {\bf 78}, 020101(R) (2008).
\bibitem{Tanaka}
D. Tanaka, Phys. Rev. Lett.  {\bf 99}, 134103 (2007).



\bibitem{Koga}
 S. Koga and Y. Kuramoto, 
\newblock  Prog. Theor. Phys. {\bf 63}, 106 (1980).

\bibitem{Ohta89}
 T. Ohta, M. Mimura and R. Kobayashi, 
\newblock  Physica D {\bf 34},  115 (1989).
\bibitem{Kerner}
 B. Kerner and V. V. Osipov, 
\newblock  Sov. Phys. Usp. {\bf 32}, 101 (1989).



\bibitem{Fel}
L. G. Fel, Phys. Rev. E {\bf 52}, 702 (1995).

\bibitem{Brand}
H. R. Brand, H. Pleiner and P. E. Cladis, Physica A {\bf 351}, 189 (2005).

\bibitem{Ohkuma09}
T. Ohkuma and T. Ohta, (to be published).

\bibitem{Sano09}
M. Y. Matsuo, Y. T. Maeda, and M. Sano, (to be published).


\end{thebibliography}
\end{document}